\input{aipcheck}

\documentclass[
    ,final            
  ]
  {aipproc}

\layoutstyle{6x9}

\begin{document}

\title{Spectral/timing evolution of black-hole binaries}

\classification{97.10.Gz, 04.25.dg}
\keywords      {Accretion and accretion disks -- black hole binaries}

\author{Tomaso M. Belloni}{
  address={INAF -- Osservatorio Astronomico di Brera, Via E. Bianchi 46, I-23807, Merate, Italy}
}

\begin{abstract}
 I briefly outline the state-paradigm that has emerged from the study of black-hole binaries with RossiXTE. This is the starting point of a number of studies that address the connection between accretion and jet ejection and the physical nature of the hard spectral components in these systems.
\end{abstract}

\maketitle


\section{States and state-transitions}

Almost fourteen years into the life of the RossiXTE mission (at the time of writing), the existing database of 
outbursts of transient black-hole binaries is very rich and constitutes a unique laboratory
to understand accretion onto black holes. The dense coverage and high-flexibility of
RossiXTE are indeed unique and not likely to be matched by new missions in the near
future. A large amount of theoretical and observational work has focused on the hard
and soft states, in particular regarding the energy spectrum (see e.g. Gilfanov 2009).
In the hard state (LHS), the energy spectrum shows a clear high-energy cutoff around $\sim$50--100 keV, while in the soft state (HSS) a cutoff was not observed up to $\sim$1 MeV (Grove et al. 1998). These spectra are interpreted as the result of Comptonization from a combination of thermal and non-thermal electrons (see e.g. Ibragimov et al. 2005). Different theoretical models differ in the details of the physical location and combination of these components.

However, a more dynamic view of the time evolution of the outbursts of these objects 
is now possible. From this, it is possible to identify a small number of states, which in 
addition to the hard and soft ones, include intermediate states. These are relatively short-lived
and mark transitions between hard and soft and vice-versa. Their importance lies in a number
of aspects. First, the study of the transition between two very different states as the hard and soft state, 
which are characterized by qualitatively different energy spectra, provides the best opportunity to 
establish their nature. In simple words: if the transition is between a thermal or hybrid Comptonization spectrum
and a non-thermal spectrum, what happens in between? Second,it has been shown that some of these
transitions correspond to changes in the infrared/radio emission (see e.g. Homan et al. 2005; Coriat et al. 2009) or to te ejection of relativistic jets (Fender, Belloni \& Gallo 2004; Fender, Homan \& Belloni 2009; Fender 2009; Gallo 2009).

The basic tools used for RossiXTE data for approaching the evolution of the X-ray emission of transient black hole binaries in outburst are the Hardness-Intensity Diagram (HID), where the total count rate is plotted as a function of hardness and the Hardness-Rms Diagram (HRD), where the fractional rms, integrated over a broad range of frequencies, is plotted versus hardness. Both diagrams are produced with X-ray photons in the approcimate 3--20 keV range. You can see several examples in Belloni et al. (2005), Homan \& Belloni (2005), Fender, Homan \& Belloni (2009), Belloni (2009), Dunn et al. (2009). A schematic picture with 
the typical path of an outburst and the states and state transitions is shown in Fig. 1. The four states are Low/Hard (LHS), Hard-Intermediate (HIMS), Soft-Intermediate (SIMS) and High-Soft (HSS) (adapted from  Belloni 2009).The vertical lines mark the state transitions. 

\begin{figure}
  \includegraphics[width=10 true cm]{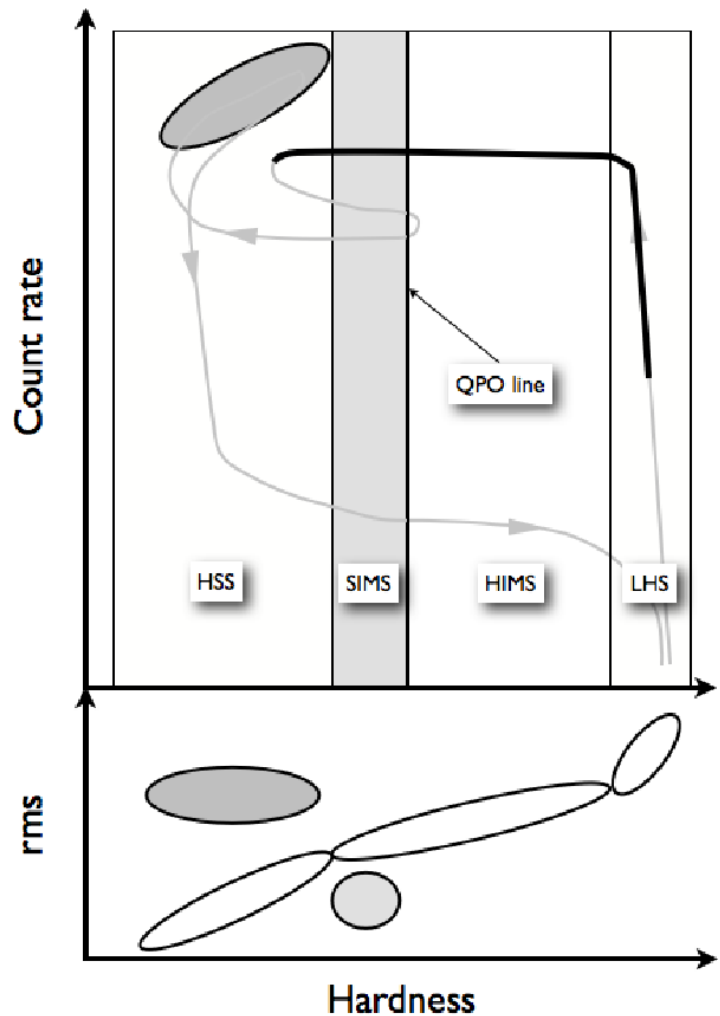}
  \caption{Top panel: sketch of the main regions of the Hardness-Intensity Diagram, corresponding
  to different sources states as identified through variability parameters. The path is an example of
  outburst evolution. The black segment of the path corresponds to the interval covered by GX 339--4
  in 2006 (see Motta, Homan \& Belloni 2009) and Fig. 2. For the definition of QPO line, see text.}
\end{figure}

State transitions are sharp: although obviously the 3--20 keV energy spectrum cannot be much different when 
crossing the vertical lines, detailed analysis of the properties of fast aperiodic variability shows that indeed
marked changes take place there (see Belloni 2009). Therefore, the position of the transition lines is not
at all arbitrary, but is fixed by observations. Particularly important is the transition between HIMS and
SIMS, marked as QPO line in Fig. 1 as it involves the appearance of a particular Quasi-Periodic Oscillation associated with the SIMS (see Belloni 2009; Casella, Belloni \& Stella 2005).

\section{Connection to jet ejection}

In the past few years, thanks to an increased observational radio/X-ray activity, a picture of the
connection between accretion and ejection is also emerging (see Fender 2009; Gallo 2009). In particular, it 
is now clear that while the LHS is associated to compact radio jets and the HSS to quenched radio emission, the major jet ejections take place during intermediate states. Fender, Belloni \& Gallo (2004) identified the QPO-line with the position corresponding to the ejection of fast jets (the ``Jet-line''). However, recently
Fender, Homan \& Belloni (2009) found that this association is not perfect nor causal: the two lines are crossed within a couple of weeks from each other, in either sequence.


\begin{figure}

  \includegraphics[width=12 true cm]{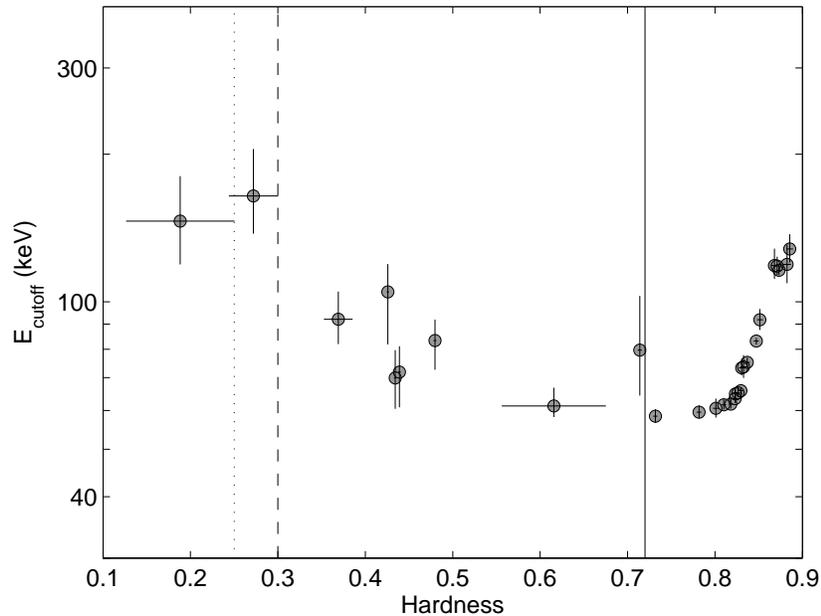}

  \caption{High-energy cutoff as a function of spectral hardness for the 2006 data of GX 339--4 (see Motta et al. 2009). The lines mark state transitions (thick: LHS-HIMS; dashed: HIMS-SIMS, the QPO-line; dotted: HIMS-SIMS).}
\end{figure}

\section{High-energy components}

Although across transitions the energy spectrum does not change appreciably, once energies above 20 keV are considered the picture changes. In particular, recent results on GX 339--4 have shown that the high-energy cutoff of the hard spectral component changes in non-monotonic way across the transition (Motta et al. 2009). Figure 2 shows the evolution of this parameter as a function of hardness, over the path sketched in black in Fig. 1. in the LHS, the high-energy cutoff decreases from 120 to 60 keV as the sources brightens (and softens), then the trend is reversed and in the HIMS there is a marked increase back to $\sim$100 keV. After the QPO-line, the cutoff energy is high, possibly not detected significantly (see also Motta et al., this volume).
This evolution is rather complex. Although the LHS behaviour is simple to understand, as the simultaneous increased soft emission cools the electrons responsible for the Comptonization. However, the trend reversal is still puzzling. The same behaviour was shown by other sources (see e.g. Joinet et al. 2008; Motta et al. 2009).

\section{Conclusions}

We now have a clear phenomenological picture of the evolution of black-hole transients, which can be compared to that of neutron-star binaries and active galactic nuclei (see Belloni 2009). It is clear that 
state-transitions hold the key to a deeper understanding. Although they are transient and difficult to observe, the data already existing show that the spectral evolution is complex and still needs to be understood.





\bibliographystyle{aipproc}   

\bibliography{sample}

\begin{thebibliography}{9}

\bibitem{1}
T.~M. Belloni, ``States and transitions in black-hole binaries,'' in \emph{The Jet Paradigm - From Microquasars to Quasars, Lect. Notes Phys. 794}, edited by T. Belloni, Springer, 2009, in press (arXiv:0909.2474)

\bibitem{2}
T.~.M. Belloni, J. Homan, P. Casella, et al., \emph{A\&A}, 2005, 440, 207 

\bibitem{3}
P. Casella, T.~M., Belloni, L. Stella, \emph{ApJ}, 2005, 629. 403

\bibitem{4}
M. Coriat, S., Corbel, M.~M. Buxton, et al., \emph{MNRAS}, 2009, 400, 123

\bibitem{5}
R. Dunn, R.~P. Fender, E. Koerding, et al., \emph{MNRAS}, 2009, in press (arXiv:0912.0142) 

\bibitem{6}
R.~P. Fender, ``'Disc-jet' coupling in black hole X-ray binaries and active galactic nuclei,'' in \emph{The Jet Paradigm - From Microquasars to Quasars, Lect. Notes Phys. 794}, edited by T. Belloni, Springer, 2009, in press (arXiv:0909.2572)

\bibitem{7}
R.~P. Fender, T.~.M. Belloni, E. Gallo, \emph{MNRAS}, 2004, 355, 1105

\bibitem{8}
R.~P. Fender, J. Homan, T.~.M. Belloni, \emph{MNRAS}, 2009, 396, 1370

\bibitem{9}
E. Gallo, ``Radio emission and jets from microquasars,'' in \emph{The Jet Paradigm - From Microquasars to Quasars, Lect. Notes Phys. 794}, edited by T. Belloni, Springer, 2009, in press (arXiv:0909.2585)

\bibitem{10}
M. Gilfanov, ``X-ray emission from black-hole binaries,'' in \emph{The Jet Paradigm - From Microquasars to Quasars, Lect. Notes Phys. 794}, edited by T. Belloni, Springer, 2009, in press (arXiv:0909.2567)

\bibitem{11}
J.~E. Grove, W.~N. Johnson, R.~A. Kroeger, et al., \emph{ApJ}, 1998, 500, 899

\bibitem{12}
J. Homan, T.~.M. Belloni, \emph{Ap\&SS}, 2005, 300, 107 

\bibitem{13}
J. Homan, M. Buxton, S. Markoff, et al., \emph{ApJ}, 2005, 624, 295

\bibitem{14}
A.~A. Ibragimov, J. Poutanen, M. Gilfanov, et al., \emph{MNRAS}, 2005, 362, 1435

\bibitem{15}
A. Joinet, E. Kalemci, F. Senziani, \emph{ApJ}, 2008, 679, 655

\bibitem{16}
S. Motta, T.~.M. Belloni, J. Homan, \emph{MNRAS}, 2004, in press (arXiv:0908.2451)


\end{thebibliography}




\end{document}